\documentclass[aps,pra,preprint,longbibliography,superscriptaddress]{revtex4-1}
\usepackage{color}
\usepackage{xcolor}
\usepackage{graphicx}
\usepackage{amsmath}
\usepackage{amsfonts}
\usepackage{multirow}

\begin{document}

\title{Broadband cantilever-enhanced photoacoustic spectroscopy in the mid-IR using supercontinuum}

\author{Tommi Mikkonen}
\affiliation{Laboratory of Photonics, Tampere University of Technology, Tampere, Finland}
\author{Caroline Amiot}
\affiliation{Laboratory of Photonics, Tampere University of Technology, Tampere, Finland}
\affiliation{Institut FEMTO-ST, UMR 6174 CNRS, Universit\'e de Bourgogne Franche-Comt\'e, Besan\c con, France} 
\author{Antti Aalto}
\affiliation{Laboratory of Photonics, Tampere University of Technology, Tampere, Finland}
\author{Kim Patokoski}
\affiliation{Laboratory of Photonics, Tampere University of Technology, Tampere, Finland}
\author{Go\"ery Genty}
\affiliation{Laboratory of Photonics, Tampere University of Technology, Tampere, Finland}
\author{Juha Toivonen}
\affiliation{Laboratory of Photonics, Tampere University of Technology, Tampere, Finland}

\begin{abstract}
We demonstrate cantilever-enhanced photoacoustic spectroscopy in the mid-infrared using a supercontinuum source. The approach is broadband, compact, and allows for higher photoacoustic signal intensity and enhanced signal-to-noise ratio as compared to systems employing conventional back body radiation sources.  Using this technique, we perform spectroscopic measurements of the full ro-vibrational band structure of water vapor at 1900 nm and methane at 3300 nm with relative signal enhancement factors of 70 and 19, respectively, when compared to measurements that use a black body radiation source. Our results offer novel perspective for photoacoustic detection opening the door to compact and sensitive broadband analyzers in the mid-infrared spectral region.
\end{abstract}

\maketitle

\section{Introduction}

Photoacoustic spectroscopy (PAS) is an optical sensing technique that detects the pressure wave resulting from local heating and thermal expansion when light is absorbed by a gas sample placed inside an acoustic cell.  Photoacoustic spectroscopy with high-power lasers is particularly attractive as the signal detected by a pressure-sensitive detector is directly proportional to the absorbed light power, resulting in highly sensitive background-free measurements. Another significant advantage of PAS is the small volume of gas sample required which makes it generally more compact than e.g. conventional Fourier transform infrared spectrometers (FTIRs). Microphones with electrical readout are commonly used as detectors in PAS, however their sensitivity is limited by the electrical noise. Resonant acoustic cells are therefore often needed for high sensitivity measurements, which limits the detection to a single frequency and absorption line. By replacing the microphone with a micromechanical cantilever whose mechanical oscillations are detected with an optical interferometer, acoustic resonance enhancement is not required, allowing for broadband detection with similar sensitivities \cite{Kauppinen:04} and indeed sub-ppt sensitivities have been reported using such a cantilever enhanced photoacoustic spectroscopy approach (CEPAS) \cite{Tomberg:18}.  

Various types of light sources have been used in laser based PAS systems, including distributed feedback diode lasers \cite{Laurila:05,Kosterev:04} and external cavity diode lasers \cite{Veres03} in the near-infrared, or quantum cascade laser \cite{Spagnolo:12,Hirschmann2:13} and optical parametric oscillators \cite{Peltola:13} in the mid-infrared. These sources are inherently narrowband or they have a limited wavelength-tuning range. Broadband detection may be achieved using thermal emitters or black body radiators but their low brightness limits the sensitivity in this case. Supercontinuum (SC) sources on the other hand can exhibit extremely broad bandwidth with brightness exceeding by orders of magnitude that of thermal emitters and the unique properties of SC sources have made them ideal light sources candidates for many sensing and imaging applications including spectroscopy, microscopy or optical coherence tomography \cite{Wildanger:08,Amiot:17,Sych:10,Hartl:01,Humbert:06}. In this work, combining a broadband supercontinuum source with FTIR CEPAS we demonstrate sensitive, broadband photoacoustic detection of water vapor and methane in the mid-infrared spectral region. Our results show a significant increase in the signal intensity as compared to when using a thermal emitter and open up new perspectives for gas sensing and more generally spectroscopic applications using broadband photoacoustic detection.

\section{Experiments}
\label{sec:examples}

\begin{figure}[tp]
\centering
\includegraphics[scale=0.8]{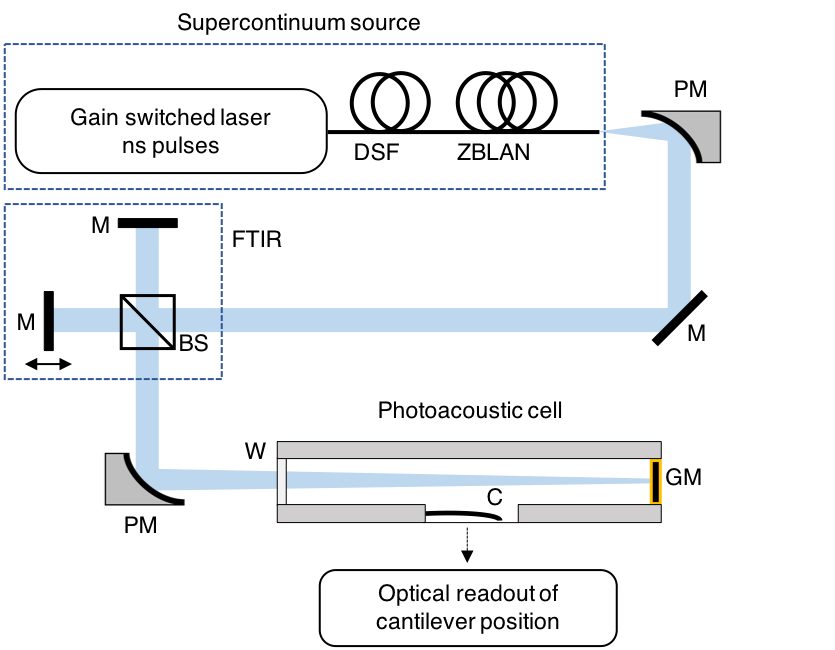}
\caption{Experimental setup. PM: off-axis parabolic mirror, M: mirror, BS: beam splitter, W: window, GM: gold mirror, C: cantilever, PD: photodiode, DL: diode laser.}
\label{fig:Setup}
\end{figure}

\begin{figure}[htp]
\centering
\includegraphics[scale=0.5]{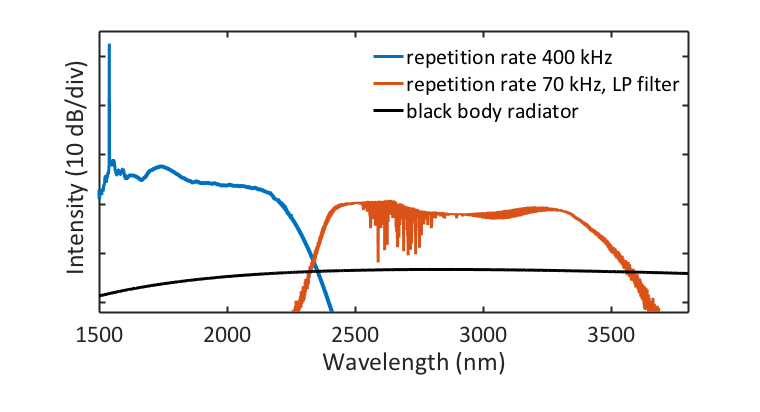}
\caption{Supercontinuum spectra corresponding to pump laser repetition rates of 70~kHz (red line) and 400~kHz (blue line) and used for the photoacoustic spectoscopic measurements of methane and water vapor, respectively. Note that the SC spectrum corresponding to 70~kHz has been spectrally filtered with a long-pass filter with cut-on wavelength of 2300~nm. The spectrum of the black body radiation source used for comparison is also shown (black line).}
\label{fig:source spectra}
\end{figure}

Our experimental setup is shown in Fig.~\ref{fig:Setup}.The supercontinuum source is generated by injecting 0.6~ns pulses at 1547 nm from a gain-switched fiber laser (Keopsys-PEFL-K09) ) with tunable repetition rate into the anomalous dispersion regime of a 4-m-long silica dispersion-shifted fiber (DSF, Corning Inc LEAF) followed  by an 8~m long fluoride fiber (ZBLAN,  Fiberlabs-ZSF-9/125-N) as described in \cite{Amiot:17}. Here, depending on the gas sample to be characterized, we used different repetition rate for the pump laser so as to optimize the power spectral density of the SC in the spectral range of interest. Specifically, for a repetition rate of 70~kHz, the 10~kW peak power of the pump laser pulses leads to the generation of a broadband SC extending from c.a. 800 nm up to c.a. 3700 nm as the result of multiple cascaded nonlinear dynamics including modulation instability, soliton formation and Raman self-frequency shift \cite{Dudley:06}. When the repetition rate is increased to 400~kHz, the peak power of the pump pulses is reduced to less than 2~kW and the spectrum extends from 1300 nm up to 2400 nm. The average SC spectra measured with a scanning monochromator and corresponding to 70~kHz and 400~kHz repetition rates for the pump laser are shown in Fig.~\ref{fig:source spectra} as the blue and red solid lines, respectively. Note that in the 70~kHz repetition rate case, the short wavelengths side of the SC spectrum was limited using a long-pass filter (Northumbria Optical Coating, SLWP-2337) with a cut-on wavelength of 2300~nm. Note that the sharp features observed at around 2600 nm in the average spectrum obtained for a 70~kHz repetition rate are caused by water vapor absorption present in air.

Light from the SC is collimated into a 2~mm (diameter) beam using a silver reflective collimator and sent to a scanning FTIR interferometer (Bruker IRCube) modulating the spectrum of the light source. At the FTIR output the beam is focused with a parabolic mirror ($f=76.2$ mm) into a non-resonant photoacoustic cell (Gasera) that contains the gas sample to be measured. The cell is gold-coated, 10 cm long and 4.5 mm diameter corresponding to a sample volume smaller than 8 mL. The front window of the cell is made of BaF$_2$ while the back window is gold coated enabling to increase the optical path length by a factor of two. The temperature and pressure of the cell were set to 50 $^\text{o}$C and 1~bar, respectively. The pressure changes inside the cell caused by light absorption in the gas sample are read optically via interferometric detection where a cantilever acts as a moving mirror whose mechanical oscillations change the period of the interference fringes. The cantilever is gold-coated with dimensions of 6 mm, 10 $\mu$m, and 1.5 mm (length, thickness, and width). The FTIR scanning mirror velocity is 1.6~kHz (monitored by a HeNe laser), corresponding to modulation frequencies of the order of few hundreds Hz for the mid-infrared wavelength range considered here and it is also the optimal frequency range for the cantilever.  The gas sample absorption spectrum is obtained by Fourier transform of the recorded interferogram. The wavelength resolution of the instrument is determined by the FTIR maximum optical path difference of 1 cm.

\section{Results and Discussion}

\begin{figure}[htp]
\centering
\includegraphics[scale=0.5]{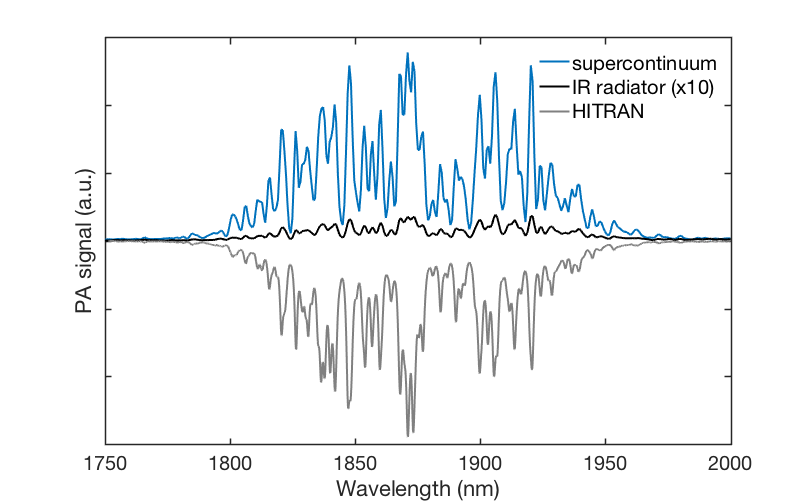}
\caption{Experimentally measured absorption spectrum of water vapor with the supercontinuum (blue line) and black body source (black line). The measurement resolution is $4 \rm cm^{-1}$ and the spectrum averaged over 10 different scans. The theoretical absorption spectrum as predicted from the HITRAN database is shown as the mirror image (grey line). Note that the results obtained with the black body have been magnified by 10 for convenient visualization.}
\label{fig:Results_H2O}
\end{figure}

The experimentally recorded absorption spectrum of normal room air (20 $^\text{o}$C, RH 30 \%) water vapor with 7000 ppm concentration in the PAS cell and using the SC source with a repetition of 400~kHz as described above is shown in Fig.~\ref{fig:Results_H2O} (solid blue line). The total SC average power was 414 mW with a power spectral density of 275 $\mu$W/nm in the water vapor absorption band. The spectrum was measured with 4 cm$^{-1}$ resolution and averaged over 10 scans for a total measurement time of 50 s. We observe excellent correspondence with the theoretical spectrum predicted from the HITRAN database (plotted as a mirror image in the figure). For comparison, we also repeated the experiment using a black body radiation source (see black solid line in Fig. 2 for an illustration of the spectrum). The measured absorption spectrum is superimposed in the figure as the black solid line. Note that the photoacoustic signal using the black body source has been magnified 10 times in the figure for visualization purposes, showing that the SC yields far better performance in terms of sensitivity and signal-to-noise ratio (SNR). More specifically, the signal intensity and SNR are increased by a factor of 70 and 13, respectively, when using the SC source. 

\begin{figure}[htp!]
\centering
\includegraphics[scale=0.5]{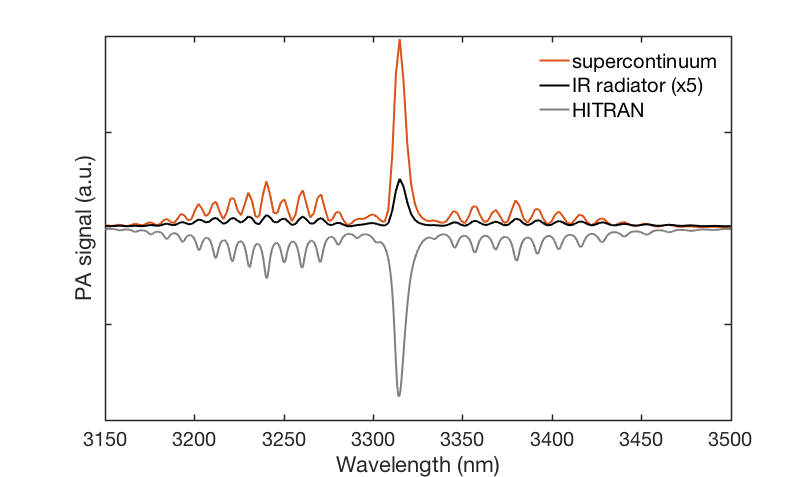}
\caption{Experimentally measured absorption spectrum of 400~ppm methane with the supercontinuum (red line) and black body source (black line). The measurement resolution is $4 \rm cm^{-1}$ and the spectrum averaged over 10 different scans. The theoretical absorption spectrum as predicted from the HITRAN database is shown as the mirror image (grey line). Note that the results obtained with the black body have been magnified by 5 for convenient visualization.}
\label{fig:Results_CH4}
\end{figure}

We then used the spectrally filtered SC source with 70~kHz repetition rate to measure the mid-infrared ro-vibrational absorption band of methane sampled from a flow of premixed 400 ppm methane in nitrogen carrier gas. The SC total power in this case (after the filter) was 78 mW with a power spectral density of 71 $\mu$W/nm in the absorption band of methane. The results are shown in Fig.~\ref{fig:Results_CH4} (red solid line) and again we find excellent agreement with the absorption spectrum predicted from HITRAN. The results from the black body radiation source are also shown in the figure (with a 5 times magnification). As in the case of the water vapor, it is clear that the signal measured when using the SC is significantly increased (19 times) as compared to that obtained with the black body source. However, the increase in the SNR is only 1.8 times in this case, which is due to increased noise in the SC source at reduced repetition rate. Indeed, there are significant pulse-to-pulse fluctuations in the SC spectrum caused by the initial stage of modulation instability in the SC generating nonlinear dynamics \cite{Dudley:06} and a lower repetition rate results in an increase in the noise level as the intensity recorded is then averaged over a  reduced number of SC pulses for lower repetition rates.   

\begin{figure}[htp!]
\centering
\includegraphics[scale=0.5]{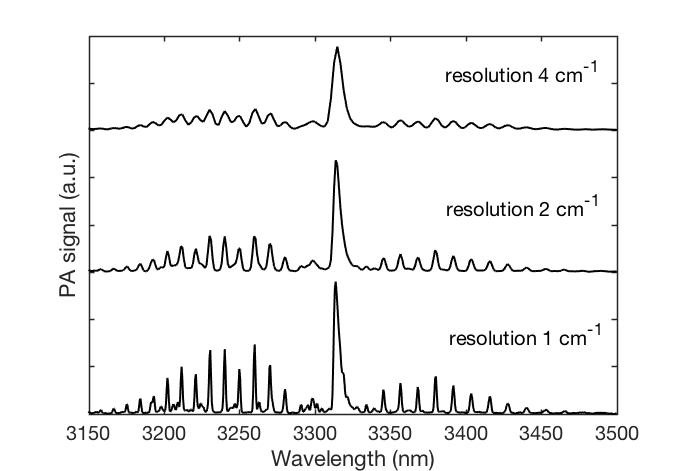}
\caption{Measured absorption spectrum of methane for different FTIR resolutions as indicated. The measurement time is inversely proportional to the resolution.}
\label{fig:Resolution}
\end{figure}

Besides the increase in the detected signal strength when using the SC source, another significant advantage as compared to a standard black body source lies in the fact that the SC is perfectly spatially coherent and the beam remains collimated independently of the optical path difference used in the FTIR. This means that one can increase the measurement resolution without any line-broadening or shifting effects \cite{Wang:17} and still maintain high signal intensity.  This is illustrated in Fig.~\ref{fig:Resolution} where we show the measured absorption spectrum of methane for increasing resolutions. One can see how the lines become sharper and the signal amplitude is increased for higher resolutions. This contrasts with the use of a spatially incoherent black body source which requires reducing the input aperture size to ensure reasonable collimation over larger distances, thus leading to decreased signal intensities when performing measurements at higher resolutions. 

We evaluated the concentration detection limits (3$\sigma$, 50 s) to be 2.6~ppm and 1.4~ppm for the black body and SC sources, respectively. The noise level was estimated by taking an average of 10 standard deviations of 10 consecutive data point blocks at the non-absorbing part of the spectral data. Lower detection limits can be achieved using resonant acoustic cells and single frequency detection or with large measurement setups. Yet, one should consider the detection limits achieved here in the context of  broadband measurement from a small sample volume. It is also evident that improved SC stability and larger spectral power density would lead to lower detection limits. 

\section{Conclusion}
We have performed broadband cantilever-enhanced photoacoustic spectroscopy using a supercontinuum light source and demonstrated the potential of the technique by measuring the absorption spectrum of water vapor and methane in the mid-infrared spectral region. The approach allows for significant improvement in terms of sensitivity and resolution as compared to the use of a conventional black body radiation source. The spatially coherent and collimated SC beam also enables miniaturization of the Fourier-transform spectrometer itself as the narrow beam can easily be guided even in a small-form-factor instrument. Furthermore, high spatial coherence enables in principle multi-pass arrangements through the PAS cell, which could lead to enhanced sensitivity. Our results open up novel perspectives for the development of cost-effective, small-footprint apparatus for broadband gas sensing. 

\section{Acknowledgments}
JT and GG acknowledge the support from the Academy of Finland (grant No. 314363). CA acknowledges the support from TUT and SPIM graduate schools.

\noindent
\bibliographystyle{naturemag}

\end{document}